# SparkGOR: A unified framework for genomic data analysis.

Sigmar K. Stefánsson[1] and Hákon Guðbjartsson[1,*]

Genuity Science, Katrínartún 4, 104 Reykjavík, Iceland

**ABSTRACT**

**Motivation:** Our goal was to combine the capabilities of Spark and GOR into a single computing framework for use in analysis of large scale genome data.
**Results:** We have created a relational query engine that unites SparkSQL and GORpipe into a single declarative query framework. This has been achieved by allowing embedding of SQL expressions into the high-level relational statement syntax in GOR and by supporting virtual relations and nested GORpipe expressions within SQL. Furthermore, we have built drivers to enable Spark and GOR to use and leverage their preferred file formats, Parquet and GORZ respectively, and introduced APIs to allow the use of GOR with Spark dataframes.
**Availability:** The SparkGOR version of the GORpipe software is open-source and freely available at https://gorpipe-website.now.sh and https://github.com/gorpipe.
**Contact:** hakon@genuitysci.com

## 1 INTRODUCTION

Like many other domains, genetics has for a while been a field of big data analysis, because of the general availability of high-throughput sequencing technology for data generation and because of continuous improvements in data storage and computing capabilities for data analysis. As data generation keeps growing at an increasing rate and there is a desire to analyze ever larger cohorts of samples, it continues to be important to improve and optimize the software analysis systems.

We released the GORpipe system few years back with them aim to provide an efficient general purpose analysis tool based on a genome ordered relational data architecture and declarative pipe-syntax query language [7]. The GORpipe system does provide a rich set of commands to work with sequence read variants and segment based data for use in rare-disease analysis as well as large scale cohort analysis. For instance, it is being used as the database for the genomics learning system in the Children's Rare Disease Cohorts initiative (CRDC) at Boston Children's Hospital[15]. Our aim here is to open-source and integrate the GOR system into Apache Spark, a popular unified general purpose computing platform for big data analysis [19].

The initial Spark architecture is based on resilient distributed data sets (RDDs) that have APIs that lend themselves elegantly to functional style programming. Because of its memory based caching architecture, it has been shown to outperform Hadoop by orders of magnitude, especially in iterative machine learning algorithms and interactive analysis. Although the general RDDs can be used to process semi-structured data, as in the MapReduce paradigm [3], much of the focus in past years has been around structured data, loosely typed relational DataFrames and DataSets with strictly typed object bindings, and SparkSQL which is arguably one of the most important and powerful feature in Spark. While the fundamental architecture of Spark has changed little over the years, the implementation has undergone significant changes between major releases. As an example are feature improvements introduced with project Tungsten, which was driven by focus on structured data and the fact that CPU had become more of a bottleneck than IO and network bandwidth. These features include off-heap memory management for reducing JVM garbage collection overhead, improvements in object serialization, cache aware data structures, code generation, and vectorized computation, leveraging columnar data formats and low level parallel instruction sets.

There are many similarities between the two platforms, GOR and Spark. Both use distributed JVM execution engine for running parallel declarative queries, both have the capabilities to access read only tabular data from various file formats, and both have a joint focus on interactive and batch processing analysis. There are also significant differences; unlike Spark that implements off-heap memory based serialization for caching of data, GOR relies on the file system client caching mechanism (e.g. NFS client caching) but caches only meta-data explicitly, such as file partition and genome positional information. The GORpipe query language supports views and syntax to refer to federated virtual relations, that may represent federated or client based data. The GOR system automatically re-evaluates such virtual relations and materializes them to a file cache using logic based on data time-stamps and expression signatures. Spark however requires explicit registration of temp view (e.g. df.createOrReplaceTempView) but uses lazy evaluation and optimizes transparently task scheduling, taking into consideration optimal memory data caching. Spark supports both SQL query style usage as well as SDK-API for functional and imperative style programming, in an interactive shell or in batch mode. Sofar, the GORpipe system has only support for multi-step pipe-query analysis and we have not yet provided APIs aimed for SDK bindings, something we change with the Spark integration presented in this work.

The primary difference between these two systems is however their standard execution approach for joins and parallel processing. GOR relies on ordered data structures and tries to optimize the use of random access vs full-scan access in merge-hash joins. For controlling parallelism it uses genome coordinates and seeks to split tasks along the genome axis, while it can also parallelize tasks along the data partition axis[1]. Results from different query steps and tasks are communicated through a shared file system. Currently, for both types of parallelism, it must be instructed explicitly through the syntax, e.g. via PGOR or PARALLEL commands. Spark however relies primarily on full-scan of data and uses broadcast or shuffle based hash-joins or sort-merge-joins (which typically also uses shuffle operations), analogous to how many SQL engines have been implemented in the past, such as those who are Hadoop based. SparkSQL also has Adaptive Query Execution (AQE), an optimization technique that uses runtime statistics to choose the most efficient query execution plan. Users can also control the level of parallelism in a more fine grained manner by using hints in SQL or through the SDK-API.

---

*Correspondence should be addressed to H.G. [1] The authors contributed equally to this work.

[1] E.g. samples for genotypes and phenotypes for GWAS results.

Several projects have used or extended Spark to work with genome data, such as ADAM [10], GATK4 [12], GenAp [9], Hail [17] and Glow [16]. The early versions of ADAM focus on secondary analysis and the use of well formed AVRO schema and Parquet columnar storage to represent sequence read and variation data. Later, they have also extended the Parquet format with metadata to better support efficient range lookups and implemented broadcast and shuffle range join logic. In GenAp, a RANGEJOIN extension to SparkSQL is introduced, similar to the high level syntax in GQL [8] and joins of pipe expressions in GORpipe. They use interval tree approach to enhance a broadcast join, to speed up the interval joins of genome features, whereas GORpipe relies on pre-ordered data and seek-scan merge join logic. In Hail the focus is on cohort analysis; kind of a Spark driven Plink system [14]. They recognize that sample based row-level representation of genotype data can create large overhead in statistical cohort analysis. They introduce Matrix tables that are similar in nature to annotated BGEN and pVCF files, implementing methods for lazy distributive analysis in Spark SDK. Glow focuses also on distributed statistical cohort analysis, however, they use standard Spark DataFrames to represent cohort genotypes, storing them in regular array-based value column that Parquet supports. This is analogous to the approach used in GOR for representing horizontal layouts, although the commands in GOR also allow data such as genotypes to be split into multiple buckets, each bucket using a separate value row for each variant. This is similar to the approach used in GenomeDB [2][13], allowing for easier updates and partitioned read access. GORpipe does also have commands that efficiently use such bucketized horizontal value format representation, for Fisher-exact test, logistic regression, LD calculation, relationship analysis, and subsetting or transposing the data back to row-level representation.

Here we present SparkGOR, our work on integrating GOR with the Spark architecture, including the following items:

- We present a relational query engine (RQE) that integrates SparkSQL and the GORpipe syntax into a generic relational query language.
- Enable parallel execution of GOR queries within the distributed Spark execution environment by building Spark-RDD query handlers for GOR.
- Extend the DataFrame API with methods for GOR, allowing the GOR expressions to be used with Spark SDK.
- Develop drivers and adapters allowing Spark and GOR to leverage each others preferred data formats and partition strategies, Parquet and GORD/GORZ respectively.

In the following sections, we describe the RQE, the unified relational query language, by showing selected examples that explain its nature and the motivation for our integration. Then we present the SDK integration with few simple Scala code snippets that show how GOR can be used with Spark DataFrames as well as an extract from a more extensive PySpark script based on GloWGR [11], that is an excellent candidate for an analysis workflow deployed with Spark Kubernetes Operator.

## 2 RELATIONAL QUERY ENGINE

To understand the extension of GORpipe to RQE requires a basic knowledge of the GORpipe syntax described in [7] and [4]. The formal structure is zero or more `CREATE` statements followed by a single `GOR` or `NOR` expression that returns the final relation. This relation can be based on one or more relations defined with the previous create steps. The reference to the create statements forms a directed acyclic graph which determines the execution order and re-evaluation dependencies. In standard GORpipe syntax the expressions start with `GOR` or `NOR` to determine ordered or non-ordered execution context. Here, in SparGOR, we add expressions that start with a `SELECT` command, indicating SparkSQL context and the Spark execution engine[2]. Unlike in regular SparkSQL expressions, that query relational data sources and files that are registered into the `sqlContext`, we additionally allow these data sources to be materialized virtual relations (`[...]`), nested GOR/NOR expressions `<(...)`, and a *tail-expression* separated with pipes, e.g. `| pipe-step(s)`.

As in previous work on GORpipe, rather than providing a formal syntax specification, we believe it is more informative to explain this with selected examples. This is especially so, because RQE does not rely on a formal syntax specification for each sub-query engine. Rather, it carves out the virtual relations and nested expressions and injects the necessary data sources appropriately. The benefits of this injection approach is that it can for instance leverage all the context specific user-defined SQL functions and is also insensitive to changes in the SQL syntax specification.

To illustrate the above, first consider the following simple case:

**Example 1:** Create statement based on SparkSQL

```
create #r1# = select * from genes.gorz
              where chrom = 'chr1';
nor [#r1#]
```

There are few things to notice here. First, that in the SQL expression the filename `genes.gorz` can be cited directly, like in GOR and NOR expressions. Second, we have built drivers to allow the SparkSQL engine to processes GORD and the block compressed GORZ file format, with seek option based on predicate push-down from genome range filter, as shown above. Thus, the read execution is quite similar as in `gor -p chr1 genes.gorz`, however, unlike the materialized outputs of GOR and NOR, that are stored in GORZ and TSV respectively, the output of a SpakSQL query, such as `[#r1#]`, is in the Parquet format which is now seamlessly read by the NOR command.

Now consider a case where a user wants to identify variants based on some rsIDs of choice, e.g. to find annotation information on allele frequencies or to locate GWAS hits. Currently, this is a use-case where GORpipe would have to use full-scan, e.g. on a narrow table that includes variants with all rsIDs. We might for instance have the file `dbsnp.gorz` with four columns (chrom,pos,ref,alt,rsIDs), i.e. a standard four column VCF variant notation and an extra column with a comma separated list of rsIDs.

---

[2] We have also designed a more explicit and generic syntax that supports loosely coupled federated execution engines, as compared to the implicit and tight integration between Spark and GOR as described here. Description of it is however outside the scope of this work.

**Example 2:** Identifying variants based on rsIDs using regular GOR

```
create #myrssnps#
     = pgor dbsnp.gorz | split rsIDs
       | rename rsIDs rsID | where rsID ~ 'rs222*';

create #myphewas#
     = pgor [#mysnps#] | varjoin -l -r PheWas.gord;

nor [#myphewas#] | sort -c pval:n,rsID
```

On an 8 core MacBook Pro computer, the `#myrssnps#` query runs in about two and a half minute, returning almost 12 thousand variants from a datasource with close to 700 million rows. Most of the time is spent reading and parsing the file. For comparison, a query that does not use a wildcard and filters on a single variant is still around one and a half minute. Although higher level of parallelization with more CPU power can be used to run this query significantly faster, it is clear that this type of full-scan execution approach uses significant amount of resources. The second stage, `#myphewas#`, does however benefit from the highly efficient seek-scan merge-join logic in GOR and runs in less than 5 sec, even though `PheWas.gord` represents GWAS hits for over 1400 PheCode phenotypes and over 140 billion rows in total.

With SparkGOR, we can now leverage SQL for the rsID lookup and write the query in the following manner:

**Example 3:** Identifying variants based on rsIDs with SparkGOR

```
create #orddbsnp#
    = select * from <(pgor dbsnp.gorz | split rsIDs
                       | rename rsIDs rsID)
       order by rsID;

create #myrssnps#
    = select * from [#orddbsnp#] where rsID like 'rs222%';

create #myordrssnps# = gor [#myrssnps#] | sort genome;

create #myphewas#
    = pgor [#myordrssnps#] | varjoin -l -r PheWas.gord;

nor [#myphewas#] | sort -c pval:n,rsID
```

Worth noting, in the `#orddbsnp#` SQL query statement, is the use of a datasource based on a nested GOR query expression. Unlike in the typical execution model in GORpipe, where `PGOR` expression is converted into multiple statements that are executed and materialized in parallel into a GORD dictionary structure, within SQL the `PGOR` expression is turned into a lazy partitioned RDD datasource that is substituted into the SQL expression. Notice that within the nested expression, we apply the `SPLIT` command that may generate multiple rows from a single row, if a variant has multiple rsIDs. This functionality is not available in regular SQL, however, analogous to a `flatMap` transformation available on DataFrames in Spark.

The SparkSQL engine will then execute the query in parallel using its preferred strategy. Sorting 700 million rows may take a while, however, in the context of our example this is a one-time execution, since it is automatically cached to files by GORpipe and is independent from the rsIDs that are of interest to the end-user (e.g. rsIDs starting with 'rs222'). Only if the file `dbsnp.gorz` changes will it be re-evaluated.

Importantly, the query behind `#myrssnps#` runs in sub-second time because the predicate push-down logic in SparkSQL can take advantage of columnar range metadata associated with row groups in the Parquet file format. Another important thing worth noting in Ex. (3) is the `#myordrssnps#` statement and the need for ordering the data. The `PGOR` and the `VARJOIN` commands in `#myphewas#` require their data sources to be ordered. Rather than performing the ordering in GOR with `SORT genome`, we could have added an `ORDER BY` command into the SQL statement that does the rsID filtering, in order to ensure that the Parquet file is ordered. Similarly, we could have pre-generated an ordered Parquet file and written Ex. 3 in the following manner:

**Example 4:** Alternative version of Example 3

```
create #myordrssnps#
    = select * from dbsnp.rsOrd.parquet
        where rsID like 'rs222%' order by chrom, pos;

create #myphewas#
    = pgor [#myordrssnps#] | varjoin -l -r PheWas.gord;

nor [#myphewas#] | sort -c pval:n,rsID
```

In the above formulation, the `PGOR` command will seamlessly add chromosome range filters on the Parquet partition files stored in `#myordrssnps#` and generate an ordered merge to preserve the genome order of the input to the `VARJOIN` command. The number of partitions will be determined by the SparkSQL logic. It is not trivial to compare the performance gains from the parallel sort in Ex. 4 because range based access of `#myordrssnps#` is slightly faster for GORZ files, which is the output in Ex. 3. For most practical cases, the speed difference will be minor. Finally, it is worth noting that it is important to sort the dbsnp Parquet file based on rsIDs in order to achieve a lookup time in a second or less. As an example, if the 700 million row file is ordered by genome location, the rsID filtering typically takes between 15 and 90 seconds, depending on cluster size and the caching state of the table.

Before we end this section, we want to give a short example using a tail-expression that can be used to bundle functionality only available in GOR into SparkSQL:

**Example 5:** DAG based filter of phenotypes

```
create #mySubjects#
 = select PID,hpo_code from phenotypes.parquet
    where year(date) > 2000
    order by PID
  | where hpo_code indag(hpo_parent_child.tsv,'HP:0001507')
  | group -gc PID -lis -sc hpo_code
  | rename lis_hpo_code HPO_codes;

gor exons.gorz | where gene_symbol = 'BRCA2'
| join -segsnp -f 20 -ir
       <(gor variants.gord -ff [#mySubjects#])
```

In the above example, we filter HPO based phenotypes using the HPO ontology DAG, stored in `hpo_parent_child.tsv`. This requires a SQL flavor with connect-by or recursive SQL, neither of which is available in SparkSQL today. A special `INDAG` operator is however available in GOR, allowing us to filter on the HPO term HP:0001507 or any narrower definition. The tail GOR commands are injected into the SQL dataframe output as a `map/flatMap/mapPartitions`[3] operation, hence executed in parallel by the SparkSQL engine. Here we are also applying a `GROUP` command, to list the HPO codes in one row per subject, thus we must be careful and order the results by `PID`. This is to ensure that the GOR grouping command "sees" all the phenotypes for each subject when the SparkSQL engine distributes the query in parallel. Currently, this is not automatically handled by our injection logic.

Alternatively, we can also use a special `DAGMAP` command and expand our high-level HPO term of interest into all the leave nodes, the `dag_node` column, and generate a small relation `#myHPOs#` with all the HPO codes, that can then be joined with the phenotype table using SQL as shown below:

**Example 6:** Alternavite version of DAG based filter

```
create #myHPOs#
 = nor hpo_parent_child.tsv | where child in ('HP:0001507')
 | dagmap -c child hpo_parent_child.tsv
 | select dag_node | distinct
 | rename dag_node code;

create #mySubjects#
 = select pheno.PID as PID,
   concat_ws(',',collect_list(hpo.code)) as HPO_codes
   from phenotypes.parquet pheno
   join [#myHPOs#] hpo
   on pheno.hpo_code = hpo.code
   where year(date) > 2000;

gor exons.gorz |... as before
```

The above formulation of the query is fully equivalent to Ex. 5, however, pushes more work into the SparkSQL engine, taking greater advantage of parallelism and predicate push-down logic.

The above shows the flexibility of being able to mix seamlessly together GORpipe queries with SparkSQL, taking advantage of the strengths of each language, execution model and data formats.

## 3 SPARKGOR SDK

Relational query languages, like SQL and the GORpipe syntax, provide an elegant declarative approach to analyse data in an efficient way. However, there are limits to that approach and it may for instance be impossible to express some iterative conditional analysis, as is possible with imperative programming languages. Interactive shells for languages like Scala and Python, integrated into notebook environments like Jupiter or Zeppelin, provide a very elegant framework for distributed data analysts with Spark through the use of RDDs and DataFrames. This allows users to mix together SQL expressions with functional and imperative programming style, enabling them to extend the standard capabilities, taking advantage of lambda functions, closures and loops. This also enables the definition of complex multi-stage workflows that return multiple outputs, something which is often difficult or impossible to express in a single GORpipe query script, thus requiring embedding GORpipe into something like bash-shell or Nextflow scripts.

In SparkGOR, we provide a library that primarily does two things: provide a constructor for SparGOR session that can generate Spark DataFrames from GORpipe expressions, and implicit Scala methods to apply tail-expressions on generic DataFrames. Again, we think this is best explained with few simple examples:

**Example 7:** Basic SDK usage in Scala

```
import org.gorpipe.spark.SparkGOR
import spark.implicits._

val sgs = SparkGOR.createSession(spark,"gorconfig.txt")

val myGenes = List('BRCA1','BRCA2').toDF("gene")

myGenes.createOrReplaceTempView("myGenes")

sgs.setCreateAndDefs("create #mygenes# = select gene
         from myGenes;
         def #exons# = exons.gorz;
         def #dbsnp# = dbsnp.gorz")

sgs.setCreate("#myexons#",
         "gor #exons#
         | inset -c gene_symbol [#mygenes#]")

val exonSnps : DataFrame = sgs.dataFrame("pgor [#myexons#]
         | join -segsnp -ir #dbsnp#
         | join -snpseg -r #genes#")

val snpCount = exonSnps.groupby("gene_symbol").count()
```

The above example demonstrates how SparkGOR is added into a Scala/Spark shell with a simple import of our gorpipe JAR. A SparkGOR factory object, that is provided a reference to the spark session and GOR config file, is then used to create one or more session objects, e.g. `sgs`. In Scala, through the use of implicits, we can easily convert a list of genes to a dataframe and associate it to the SQL session as a temporary view, e.g. `myGenes`. Through the SparkGOR session object, we can then specify `CREATE` and `DEF` statements that become "visible" in our dataframe definition, as in `exonSnps`. Here we show two methods to do this; one that initializes multiple creates and definitions through one invocation and one that incrementally adds one create to the session, `sgs`.

Note that the dataframe definition is lazy and the `PGOR` command is not materialized to files, as when executed by the GORpipe engine. Rather, it generates RDD partitions[4] that are provided to the Spark engine, e.g. when the results are collected through the evaluation of `snpCount`. It might help explaining this to write the `snpCount` in alternative way:

**Example 8:** Alternative form of `snpCount`

```
val snpCount = sgs.dataFrame("
       select count(*) from <(pgor [#myexons#]
       | join -segsnp -ir #dbsnp#
       | join -snpseg -r #genes#) group by gene_symbol")
```

---

[3] Dependent on the nature of the GOR commands.

[4] `PGOR` uses a config file to specify default level of parallelization which can be overwritten with the `-SPLIT` option.

Notice, that because we are evaluating this SparkGOR expression in the context of `sgs`, our definitions for `#myexons#` and `#genes#` are available. The create statement may or may not have been materialized already. If not, it will be materialized when the lazy dataframe is evaluated.

As an example of how GOR can be used as tail-expression on a generic dataframe, based on `myVars.parquet` with variants, consider the following:

**Example 9:** DataFrames with GOR tail-expression

```
val dbsnpDf = spark.read.parquet("dbsnp.parquet")

val myVars = dbsnpDf.gor("calc type
             = if(len(ref)=len(alt),'Snp','InDel')")

myVars.createOrReplaceTempView("myVars")

sgs.setDef("#VEP#","vep.gorz")

val myVarsAnno = sgs.dataFrame("
        select * from myVars
        order by chrom, pos").gor("varnorm -left ref alt
 | group 1 -gc ref,alt,type -set -sc rsID
 | rename set_rsIDs rsIDs
 | varjoin -r -l -e 'NA' <(gor #VEP#
                          | select 1-alt,consequence)")
```

Here we define the dataframe `dbsnpDf` from a regular Parquet file. Then we use a tail-expression to add a `type` column, something which is also easily done using standard `map` functions in conventional Spark. Next we register the dataframe as temp view and add a definition into the SparkGOR session, `sgs`. This allows us to refer the data in SparkSQL and to order it as in GOR. Finally, we add add a more complex tail-expression that left-normalizes the variants, collapses rsIDs of equivalent variants into one row, followed by a left-varjoin to annotate them with VEP-consequence. It is important to recognize that both `VARNORM` and `VARJOIN` can only be executed on a dataframe that was generated from a SparGOR session, because behind the scenes these commands use a genome reference build file that is defined in `gorconfig.txt` as shown in Ex. 7. The fact that one can indeed apply GOR tail-expressions on a regular dataframe, like `dbsnpDf`, is through implicit magic in Scala, however, commands that require the genome reference context will not work. Another thing worth pointing out is that when GOR generates a dataframe, the schema is automatically inferred from the query expression to generate Spark SQL row. Similarly, GOR can use row models from SparkSQL within the GOR code base, thus there is minimal "impedance mismatch".

Next we might want to persist the results to files[5]. This can be done as show below:

**Example 10:** Writing output to files

```
myVarsAnno.write.save("myVars.parquet")

myVarsAnno.write.
   format("gorz").partitionBy("chrom").save("myVars.gord")
```

---
[5] Not to be confused with the .persist() and .cache() methods of a DataFrame that are used to cache data, to disk or memory respectively.

The former command writes the results to a regular Parquet file. Actually, the SparkSQL engine might choose to execute `myVarsAnno` in parallel, hence the output `myVars.parquet` could be a folder of Parquet files. Similarly, we can write the output in a GOR format. We have adjusted our dictionary table format to be analogous to Parquet, i.e. `myVars.gord` is a folder with GORZ files as well as a GORD file as described in [7].

Our last example shows how easily GOR can now be integrated into a whole genome regression analysis workflow, written in PySpark. The workflow uses a machine learning method called REGENIE [11], which uses linear mixed models to account for population structure and relatedness, like several other similar statistical algorithms. Unlike former approaches, they build a combined phenotype predictor based on multiple local Ridge regression models, each accounting for the correlation of local linkage disequilibrium within blocks of common variants and the phenotypes. Importantly, their block based approximation approach in the model fitting step has linear computational complexity with the number of samples. Furthermore, the block based approach makes this algorithm ideally suited to take advantage of distributed computing frameworks, such as Spark. The same applies to the second step, where the adjusted phenotype model is typically used in regression against many more variants, e.g dozens of million as compared to few hundred thousand variants in the first step.

A Spark version of this algorithm has been implemented in Glow and made available as PySpark code in a notebook [1]. We show here extracts of this PySpark code with focus on the parts that involve selection of the phenotypes and the setup of the variants dataframe:

**Example 11:** Whole genome regression with GloWGR

```
...
import spark._jvm.org.gorpipe.spark
sc = spark.sparkContext
...
label_df = pd.read_csv(phenotypes_path,
                       index_col='sample_id')

pheno_df = spark.createDataFrame(label_df)
pheno_df.createOrReplaceTempView("pheno")

sgs = SparkGOR.createSession(spark._jsparkSession)

sgs.setCreateAndDefs("
    create #pheno# = select * from pheno;
    create #samples# = nor [#pheno#]
        | inset -c sample_id varbuckets.tsv
        | select sample_id;
    create #regionsplit# = gor LDpruned_variants.gorz
        | group 1000 -count | seghist 10000")

ds1 = sgs.dataFrame("pgor -split [#regionsplit#]
     variants.gord -nf -ff [#samples#]
    | varjoin -i step1_variants.gorz
    | csvsel -u 3 -gc ref,alt -vs 1
             varbuckets.tsv [#samples#]
    | rename Chrom contigName
    | rename pos start
    | rename ref referenceAllele
    | rename alt alternateAlleles")
    .withColumn('values',chartodoublearray(col('values')))

variant1 = _java2py(sc,ds1)
```

```
...
variants1_dfm = variant1.withColumn('values',
            mean_substitute(col('values'))
            .filter(size(array_distinct('values')) > 1)
...
sample_ids = sgs.dataFrame("nor [#samples#]")

block_df, sample_blocks = glow.wgr.functions
        .block_variants_and_samples(variants1_df,
                                    sample_ids,
                                    variants_per_block,
                                    sample_block_count)
...
wgr_gwas = variants2_df
  .join(adjusted_phenotypes, ['contigName'])
  .select('contigName',
          'start',
          'names',
          'label',
          expand_struct(linear_regression_gwas(
                        variant_df.values,
                        adjusted_phenotypes.values,
                        lit(covariates.to_numpy()))) )
          )
```

As always, we need to import our SparkGOR library. From the Pandas phenotype structure, `label_df`, we create a Spark dataframe and register it as a temporary view, for it to be accessible in our SparkGOR session. As shown in the definition of `#samples#`, the samples IDs are found from the overlap of samples with phenotypes and those with genotypes, based on the variants bucket metadata file `varbuckets.tsv`. In the original notebook workflow, however, the list of sample IDs is read from the pVCF datasource header. The regions used for the parallel partitions, `#regionsplit#`, are defined such that there are 10k variants in each region, resulting in about 100 evenly sized partitions[6], if 1 million common LD-pruned variants are used for the model step. We then define our variant dataframe with GOR expression as shown with `ds1`. The GOR expression reads only the GORD file partitions that are needed based on the samples used, e.g. `-ff [#samples#]`, and a simple intersect varjoin is used to extract only the variants of interest. In the GOR horizontal genotype format, the genotypes are already bi-allelic, so a multi-allelic split, as in the original notebook, is not needed.

The `CSVSEL` command is then used to convert multiple bucket based variant rows into a single row for each variant, with a "horizontal" `values` column, for the genotypes of interest and in the order specified by `#samples#`. In GOR, `values` is a regular string typed column. Therefore, we convert it to a Spark array, using the `withColumn` DataFrame method and `chartodoublearray`, a UDF that we defined in order to format the genotype array such that it is compatible with the Glow `values` column. The definition of the variant input to step 2, `variants2_df`, is analogous to the one for step 1, except that it does not use `LDpruned_variants.gorz`, but rather a different file or no intersection varjoin step, to include all available variants.

Note that the GOR expression dataframes for the variants are lazy and not evaluated until the Glow regression steps fetch the variants, e.g. in the evaluation of `wgr_gwas` in regression step 2. Thus, there is no explicit write of intermediary temporary files here, although the execution engine may create some behind the scenes, e.g. depending on the amount of available memory for shuffle operations.

---

[6] `SEGHIST` sizes segments based on histogram equalization like approach.

Regarding the above, there is room for optimization in the above script. The Glow method `block_variants_and_samples` does indeed not assume genome ordering or partition of the dataframe based on the samples, therefore it always invokes re-partitioning and ordering of the data, to split it up into `block_df`. In GOR, the genotype data is already stored in ordered partitions and we can easily re-partitions the subset of samples efficiently into the appropriate blocks as show below:

**Example 12:** Optimized block dataframes

```
sgs.setCreate("#samples#",
            "nor [#pheno#]
              | map -c sample_id varbuckets.tsv
              | sort -c bucket,sample_id
              | select sample_id")

sgs.setCreate("#BlockRegionSplit#",
            "gor LDpruned_variants.gorz
              | group 1000 -count | seghist {:d}"
            .format(variants_per_block))

sgs.setCreate("#SampleBlocks#",
            "nor [#samples#]
              | rownum
              | calc sample_block div(rownum,{:d})
              | replace sample_id squote(sample_id)
              | group -gc sample_block -lis -sc sample_id
              | calc size listsize(lis_sample_id)"
            .format(sample_block_count))

sgs.setCreate("#RegionSampleBlocks#",
            "nor [#BlockRegionSplit#]
              | replace #2 #2+1 /* -p is one-based */
              | calc header_block #1+'_'+#2+'_'+#3
              | multimap -cartesian [#SampleBlocks#]");

block_df = sgs.dataFrame("
   parallel -parts [#RegionSampleBlocks#]
      <(gor variants.gord -nf
        -p #{col:chrom}:#{col:bpStart}-#{col:bpStop}
        -f #{col:lis_sample_id}
        | varjoin -i step1_variants.gorz
        | csvsel -u 3 -gc ref,alt -vs 1
          varbuckets.tsv <(nor [#samples#]
            | where sample_id in (#{col:lis_sample_id}))
        | calc header_block #{col:header_block}
        | calc sample_block #{col:sample_block}
        | calc size #{col:size}
        | rename ... as before)").withColumn(...as before
```

Notice that we have modified the definition of `#samples#` such that it is ordered according the bucket structure of our variant source, `variants.gord`, as specified in `varbuckets.tsv`. This is to ensure that each GORZ bucket file is read by as few partition threads as possible. Now we define `#BlockRegionSplit#` to include the numbers of variants desired for each block in the model step, as specified by the variable `variants_per_block`. Then we group the samples into blocks in `#SampleBlocks#` and finally we use a Cartesian multimap-join to setup two dimensional blocks, for the variants and the samples, i.e. `#RegionSampleBlocks#`.

The macro command `PARALLEL` is then used to create as many parallel partitions as there are rows in `#RegionSampleBlocks#`,

allowing reference to the columns of this parts relation via `#{col:colname}` to be used as input to each query task. Notice that the `-p` option specifies the region in each block and similarly does the partition filtering option, `-f`, specify the samples that are needed. However, since the horizontal bucket rows store values from multiple samples, we extract values from the rows only for the samples that belong to the block. This is done with the NOR expression, provided as the second parameter to the `CSVSEL` command. The calculation of `SIZE` and the two other block columns is to be compatible with the output from the original Glow method.

Finally, we write the regression results to disk, first as Parquet files and then as GORZ dictionary files, simply to show the possibilities.

**Example 13:** Writing GWAS output for many traits

```
wgr_gwas.createOrReplaceTempView("wgrgwas")

spark.sql("select * from wgrgwas
          order by label, chrom, pos")
     .write.format("parquet").partitionBy("label")
     .save("wgr_gwas.parquet")

sgs.dataFrame("select * from wgrgwas")
   .repartition("label")
   .gor("sort genome")
   .write.format("gorz").partitionby("label")
   .save("wgr_gwas.gord")
```

In both cases, we partition based on trait[7] and order based on genome position. In the latter case, we show an example where a GOR tail-expression is used on a dataframe, to do the ordering within each trait. In supplementary material [5], we show more examples including how GWAS association results, from close to 1500 PheCode traits defined from the UKBB data, can be interrogated using the query capabilities of SparkGOR.

On small datasets when there are few samples and variants in the model step, the speed difference between the approach shown in Ex. 11 and Ex. 12 may not play a major role and in general, the steps involving the RidgeReducer()[8] may overshadow the shuffling overhead in ordering the variants, especially when multiple trait are analyzed together. Nevertheless, the above example presents a relatively easy path to integrate GOR in an effective manner into a Glow based algorithm. Clearly, the SparkGOR SDK integration opens up multiple options and choices for working with genome data using a growing number of libraries that are built for the Spark platform, such as Glow, ADAM, GATK, and Hail.

## 4 DISCUSSION

In this paper, we have presented an integration of GOR with Spark, first as a unified relational query engine, mixing declarative SparkSQL and GORpipe queries, and secondly as a library to work with Spark dataframes, e.g. in Scala or Python. We have discussed the similarities between these two architectures as well as their fundamental differences, such as their approach for joining relations. It is our belief that each approach brings its own strengths and weaknesses, all depending on the use-cases at hand. This work was however not intended to be a formal comparison or benchmark of the two architectures.

Overall, we can however say that many use-cases based on range joins and range access are more efficient with GOR and less dependent on memory caching. For instance, the dynamically optimized seek-scan approach in GORpipe makes is possible join the entire ClinVar database, with few hundred thousand unevenly distributed variants, against a table with seven hundred million annotated variants, in about five seconds on a MacBook Pro. This takes ten times longer time in SparkSQL using the same hardware, even when using its most effective broadcast join approach. Likewise, queries that make fuzzy join on genome segments, such as exons or genes, or use a single range scan invoked from a genome browser, run very fast in GOR but are significantly less efficient in Spark. Hence, de-normalization is often needed in SparkSQL to avoid genome range joins, thus relying on columnar access and efficient filtering.

Filtering speed of rows is significantly faster in SparkSQL, especially if the data has been cached to memory or if the filter is highly selective and can take advantage of row groups metadata in the Parquet format. Nevertheless, in the latter case, the performance ends up being sensitive to the ordering strategy of the table, as highlighted in some of our examples. Indeed, the partition strategy does play a major role for both systems, especially for large data sets. Compression ratio in GORZ and Parquet is similar while the computation cost of writing is less in GORZ. Also, Spark demands a significant amount of memory in order to work well and the Parquet driver seems to have a significantly higher memory footprint per file than the GORZ driver.

There have been efforts to address some of the above issues. For instance, the ADAM project [10] has introduced a genome range partition scheme extension to Parquet for their `GenomicDataset`, allowing for a more efficient filtering by overlapping regions. In particular, they report that the time of range scans goes from few minutes to few seconds. For comparison, a typical range scan executed from a GOR based genome browser, fetching 1000 rows from a Clinvar GORZ table takes less than 100msec.

Their approach is indeed similar to our GOR dictionary files, i.e. the output of `PGOR` queries where the file partitions and corresponding genome range is stored in a GORD metadata file. We have also started to experiment with dictionaries based on Parquet files (currently coined GORP), in order to non-intrusively leverage Parquet columnar projection and predicate push-down within GOR. However, since the GOR query logic is based on seek-able ordered iterators[9], it must be done without introducing longtime blockage from iterators that use the Parquet filtering mechanism. By using genome ordered and moderately sized range partitions and by extending our iterator interface, such that iterators provide position progress information to its consumer, we have been able to adopt the Parquet driver with our execution patterns, i.e. where data is merged from multiple files and dynamic seek-scan decisions are made based on the streaming progress. Similar blockage issues come up in relation to merge patterns from federated ordered iterators, that implement predicate push-downs or more complex query logic such as grouping, patterns that can be of value to increase parallelism or reduce the impact of network bottlenecks[18].

---

[7] Here called `label` to be compatible with the original notebook.

[8] A step not shown here but in the original GloWGR notebook.

[9] Different from the standard Volcano iterator model[6]

A detailed description of the above development is outside the scope of this work and awaits publication elsewhere. Same applies to other areas of further integration that lie ahead, for instance in relation to increased usage and generation of metadata to automate GOR execution within the Spark engine. It is our hope that open-sourcing SparkGOR will spur innovation and widen the interest in the unified platform.